\documentclass{article}

\usepackage{amssymb}
\usepackage{amsmath}
\usepackage{amsthm}
\usepackage{xspace}
\usepackage{comment}
\usepackage{url}
\usepackage{color}
\usepackage{enumerate}
\usepackage{centernot}
\usepackage[numbers]{natbib}

\DeclareMathOperator*{\argmax}{arg\,max}

\def \eps {\epsilon}

\newcommand{\eqr}[1]{Eq.~\eqref{eq:#1}}

\newcommand {\set}[1]{\ensuremath{\left\{#1\right\}}}

\newcommand{\abs}[1]{|#1|}

\DeclareMathOperator*{\E}{\mathbb{E}}

\newcommand{\h}{\frac{1}{2}}

\makeatletter
\@ifundefined{theorem}{%
\newtheorem {theorem}{Theorem}}{}
\@ifundefined{lemma}{%
}{}
\@ifundefined{corollary}{%
}{}

\newcommand{\ph}{\hat{p}}
\newcommand{\pb}{\bar{p}}

\newcommand{\Q}{\mathcal{Q}}  %
\newcommand{\CS}{\mathcal{C}} %
\renewcommand{\Pr}{\text{Pr}}
\newcommand{\PQ}{\Pr^\Q}
\newcommand{\PrC}{\Pr_{\CS}}

\renewcommand{\E}{\mathbb{E}}
\newcommand{\EC}{\mathbb{E}_{\CS}}
\newcommand{\EV}{\text{EV}}
\newcommand{\mmid}{\!\mid\!}

\definecolor{darkblue}{rgb}{0,0,0.7}

\newcommand{\one}{\texttt{ONE}\xspace}
\newcommand{\all}{\texttt{ALL}\xspace}

\newcommand{\imp}{\xspace$\Longrightarrow$\xspace}
\newcommand{\nimp}{\xspace$\centernot\Longrightarrow$\xspace}

\newcommand{\ch}[1]{\multicolumn{1}{c}{#1}}

\newcommand{\Qnice}{Q1\xspace}
\newcommand{\Qpoly}{Q2\xspace}
\newcommand{\Qfix}{Q3\xspace}
\newcommand{\Qconv}{Q4\xspace}

\begin{document}

\title{On Calibrated Predictions \\
for Auction Selection Mechanisms}

\author{H. Brendan McMahan\\
\texttt{mcmahan@google.com}
\and
Omkar Muralidharan \\
\texttt{omuralidharan@google.com}
}
\date{Google, Inc.}
\maketitle

\begin{abstract}  %
  Calibration is a basic property for prediction systems, and
  algorithms for achieving it are well-studied in both statistics and
  machine learning.  In many applications, however, the predictions
  are used to make decisions that select which observations are made.
  This makes calibration difficult, as adjusting predictions to
  achieve calibration changes future data.  We focus on
  click-through-rate (CTR) prediction for search ad auctions.  Here,
  CTR predictions are used by an auction that determines which ads are
  shown, and we want to maximize the value generated by the auction.

  We show that certain natural notions of calibration can be
  impossible to achieve, depending on the details of the auction. We
  also show that it can be impossible to maximize auction efficiency
  while using calibrated predictions. Finally, we give conditions
  under which calibration is achievable and simultaneously maximizes
  auction efficiency: roughly speaking, bids and queries must not
  contain information about CTRs that is not already captured by the
  predictions.
\end{abstract}

\section{Introduction}

Calibration is a fundamental measure of accuracy in prediction
problems: if we group all the events a predictor says happen with probability
$p$, about a $p$ fraction should occur. This property has been
extensively studied in the stochastic and online settings.

We study problems where the predictions themselves partially determine
which events occur.  Our general approach applies to many problems
where predictions are used to make decisions, but we are motivated in
particular by the application to search engine advertising.  Over the
past decade, this business has grown to tens of billions of dollars,
and prediction systems play a fundamental role.  

In a typical interaction, first a user does a query (say ``flowers'')
on a search engine.  Then, the search engine selects a set of
candidate ads that can be shown on the given query, based on keywords
provided by advertisers.  These components can be reasonably
approximated by an IID process.  A prediction is made for each
candidate ad, and an auction ranks the ads based on the prediction and
the bid of the advertiser.  Typically, the bid indicates the value of
a click to the advertiser, and the score is simply the product of the
bid and the prediction, giving an estimate of the value generated by
showing the ad.  Finally, some of these ads are shown to the user (we
consider two models: the single top-ranked ad is shown, or all the ads
with scores above a certain threshold are shown).  This auction
selection mechanism has been extensively studied, and has many nice
properties~\citep{varian07auctions,edelman07internet}.

In this setting, an important measure of the quality of the
predictions is how much value the auction generates (equivalently, how
\emph{efficient} are the allocations produced by the auction).  The
auction mechanisms we consider are in fact designed to maximize the
combined value to the search engine and advertiser if bids accurately
reflect value and the true click-through-rates (CTRs) are known.

The algorithm used to predict CTRs for such a system faces many
constraints already, for example, the need to process enormous volumes
of data quickly and produce predictions with extremely low latency
(e.g., \citep{graepel10webscale}).  Thus, rather than advocating new
algorithms, we focus on applying a post-correction via a prediction
map to the outputs of an existing system in order to improve the
quality of the predictions.

We consider two main questions.  Informally stated: 1) Do
efficiency-maximizing prediction maps with calibration properties
exist, and can they can be found computationally efficiently?  2) If
we iteratively calibrate our predictions so they match observed CTRs,
does the process converge?  And if so, is this prediction map
efficiency maximizing?

\paragraph{Outline and Summary of Results}
We formalize our model and questions in
Section~\ref{sec:formalization}, where we introduce two primary
variants of the selection mechanism that lead to different properties;
Section~\ref{sec:all} and \ref{sec:one} investigate these mechanisms
in the general case.  We demonstrate that without further assumptions,
in both our models it may be impossible for a deterministic prediction
map to produce calibrated predictions on the ads it serves, and
iterative calibration procedures can fail badly.  Since some
deterministic map always maximizes value, this is unfortunate.  When
all ads above a certain threshold are shown, we give an algorithm for
finding this value-maximizing map in polynomial time, but when the
single highest-rated ad is shown, we prove finding the
value-maximizing map is NP-hard (even if we knew the true CTRs).

In Section~\ref{sec:cond} we introduce additional assumptions that are
sufficient to guarantee calibration procedures are well-behaved.
While these assumptions are fairly strong, they are not unreasonable
for real systems.  Our strongest assumption is essentially that in all
cases bid and query provide no more information than the raw
prediction about average CTRs; under this assumption, we can show in
both selection models a value-maximizing and calibrated prediction map
exists.  Under threshold selection, somewhat weaker conditions are in
fact sufficient.

\paragraph{Related Work}
Calibration has been extensively studied.  Much of the earliest work
is in the probabilistic forecasting literature
\citep{brier50verification, dawid82well, Ranjan2010}.  Calibration is
particularly important when comparing predictors, since two sets of
calibrated predictions can be fairly evaluated by how concentrated
they are on observed outcomes \citep{DeGroot1983, Gneiting2007,
  Gneiting2007a}. Calibration also makes it easier to use
predictions. For example, it is easier to threshold the output of a
calibrated classifier to minimize weighted classification error
\citep{Cohen2004}.

Not all prediction systems are naturally calibrated.  However, when
examples are drawn IID, if we have a good but uncalibrated predictor,
we can calibrate it by applying a prediction map.  For example,
boosted trees are uncalibrated, but become excellent probability
estimators after calibration \citep{Niculescu-Mizil2005, Caruana2006}.
The two most common methods for calibration are Platt scaling, which
is equivalent to logistic regression, and isotonic regression
\citep{Platt1999,Zadrozny2002,caruana05prob,caruana06compare}.

Calibration is also studied in the online setting, where no stochastic
assumptions are made on the sequence of examples; in the worst case,
they could be chosen by an adversary that sees our predictions.  It is
easy to see that in this setting, no deterministic classifier (or
prediction map) can produce calibrated predictions for all sequences.
However, if the system is allowed to use randomness (that is, predict
a distribution), then calibration can be achieved
(\citep{foster96calibration,Foster1999} and \citep[Sec
4.5]{cesabianchi06plg}).

\section{Problem Formalization}\label{sec:formalization}

The interaction of calibration and selection has received little
direct attention in the literature, so constructing a suitable model
requires some care: we require a formulation that is theoretically
tractable but still captures the key characteristics of the real-world
problems of interest.

We begin by defining our units of prediction (queries and ads) and the
mechanism used to select them (auctions).  We assume a fixed, existing
prediction system provides a raw prediction for each ad; our study
will then concern prediction maps, functions that attempt to map
these raw predictions to calibrated probabilities.  Once this
framework is established, we can formally state the questions we
study.

We model the interaction between a search engine's users and
advertising system.  There is a fixed finite set of queries $\Q$
(strings like ``flowers'' or ``car insurance'' typed into the search
engine), which are chosen according to distribution $\PQ(q)$ for $q
\in \Q$.
There is also a fixed finite set of ads $\CS$ which can be shown
alongside queries.  Each ad $i \in \CS$ is defined by tuple $(p_i,
b_i, z_i, q_i)$ where $q_i \in \Q$ is the (only) query for which ad
$i$ can show,\footnote{This is without loss of generality, as we can
  always replicate ads for each query to which the advertiser has
  targeted the ad.} $p_i$ is the true probability of a click, $b_i$ is
the bid (the maximum amount the advertiser is willing to pay for a
click), and $z_i \in \set{1, \dots, K}$ is a bucketed estimate of
$p_i$ (we call $z_i$ the raw prediction).  That is, we assume the
predictions of the underlying prediction system have been discretized
into $K$ buckets.  We drop the $q$ (and sometimes $z$) from the ad
tuples when those values are clear from context.  Each ad can show for
a single query $q$, so we define $\CS(q) \equiv \set{i \mmid q_i = q}$,
the indexes of the candidate ads for query $q$.

Our goal is to find good prediction maps $f : \set{1, \dots, K}
\rightarrow [0, 1]$.  The prediction map will be used in the auction
selection mechanism: First, a query is sampled from $\PQ$, and then
the candidate ads for that query are ranked by $b \cdot f(z)$ (we drop
the subscripts when we mean an arbitrary ad).  We consider two models
for which ads show:
\begin{description}
\item[\one:] We only show a single ad.  If multiple ads achieve the highest
  value of $b \cdot f(z)$, we pick one uniformly at random.
\item[\all:] We show all ads where $b \cdot f(z) - 1 > 0$.
\end{description}
Mechanism \one models the case of an oversold auction, where ads with
different raw predictions $z$ must compete for a single position.
Mechanism \all models the case where all eligible ads with positive
predicted value can be shown.  In general, mechanism \all is much
easier to work with theoretically, because for $z_1 \neq z_2$,
changing $f(z_1)$ does not change which ads with prediction $z_2$ are
shown.  In either case, we assume any candidate $(p, b, z)$ which is
shown is clicked with probability $p$.\footnote{This ignores the
  well-known issue of position normalization; this aspect of the
  problem is largely orthogonal to our work. 
}

\paragraph{Distributions on Ads} Other than the distribution $\PQ$,
all probabilities and expectations will be with respect to some
distribution on the set of candidate ads $\CS$.  Two distributions
will be of particular importance: $\PrC$, the uniform distribution
over candidate ads, and $\Pr_f$, the distribution of ads shown by
a prediction map $f$.  We formalize these as follows:

$\PrC$ is the distribution on ads where $\PrC(i)$ is
proportional to $\PQ(q_i)$.  That is, letting
$
C \equiv \sum_{i \in \CS} \PQ(q_i),
$
we have 
$
\PrC(i) = \frac{\PQ(q_i)}{C}.
$
This is not the same as choosing a random query $q$ from $\PQ$ and
then choosing a random candidate.  For example, suppose there are
two queries $q_1$ and $q_2$, with $\PQ(q_1) = \h$ and $\PQ(q_2) =
\h$.  There is one candidate $a_1$ for query $q_1$, and two
candidates, $a_2$ and $a_3$ for query $q_2$.  Then, $\PrC(a_i) =
1/3$ for each ad, which means the marginal probability $\PrC(q_1) =
\frac{1}{3}$ and $\PrC(q_2) = \frac{2}{3}$.  One can think of $\PrC$
as the distribution on ads shown if we showed all the eligible
candidates for each query that occurs.

$\Pr_f$ for a prediction map $f$ is the distribution on ads
where $\Pr_f(i)$ is proportional to $w_i \equiv \PQ(q_i)
\Pr(\text{ad $i$ shows} \mmid q_i, f)$.  The second term is actually
only random in the case of selection mechanism \one, when randomness is
used to break ties.  The distribution $\Pr_f$ is thus the
distribution on ads shown when serving using prediction map $f$.
Using this notation, $\Pr_f(i \mmid q) =\Pr(\text{ad $i$ shows} \mid
q_i, f)$.

We use $\EC[\cdot]$ and $\E_f[\cdot]$ for the corresponding
expectations.

\paragraph{Calibration}
We say a prediction map $f$ is calibrated on a distribution on ads $D$ if
\[ 
\forall z, \quad 
\underbrace{\E_{(p,b,z,q) \sim D}[ p \mmid z]}_{ \text{Average CTR
    given $z$}} = \underbrace{f(z).}_{ \text{Predicted CTR given $z$}}
\] 

The choice of the distribution $D$ in the above definition is
critical; a single $f$ will in general not be able to achieve
calibration for multiple $D$.  For the auction selection problem, the
natural distribution to consider is $\Pr_f$.  Thus, we will be
particularly concerned with finding \emph{self-calibrated} prediction
maps $f$, which satisfy
\[ \forall z, \quad \E_f[ p \mmid z] = f(z).\]

In general one may not be able to estimate $\E_f[\, p\! \mid\! z \,]$
exactly, and so calibration will only be approximately achievable.
This issue is orthogonal to our results, so we assume that the
necessary expected quantities can be estimated exactly.  Thus, we
emphasize that our negative results are a fundamental limitation,
rather than a byproduct of insufficient data.

\paragraph{Auction Efficiency} In addition to calibration, we are
concerned with how the choice of $f$ impacts the auction mechanism.
The expected value of showing ad $(p, b)$ is $p \cdot b -
\text{cost}$, where we take cost = 1 for selection mechanism \all, and
cost = 0 for \one.  We assume the bid $b$ reflects the true value to
the advertiser of a click, which is justified by the incentives of the
auction under a suitable pricing scheme~\cite{varian07auctions}.  The
cost can be viewed as the cost per impression of showing the ad
(either a cost incurred by the user doing the query or incurred by the
search engine itself). In practice such costs might be different for
clicked versus unclicked ad impressions, and might vary depending on
the ad and query.  Extending our results to such a models would add a
significant notational burden, so we focus on the simplest interesting
cost models.

For a given query $q$, the expected value generated is
\[ 
  \sum_{i \in C(q)} \Pr(\text{ad $i$ shows} \mmid f, q) (p_i b_i - \text{cost}).
\]
The expected value per query is just
\begin{align*}
\EV(f) &= \sum_{q \in Q} \PQ(q) \sum_{i \in C(q)} \Pr_f(i \mmid q) (p_i b_i - \text{cost})\\
          &= \sum_{i \in C} w_i (p_i b_i - \text{cost}).
\end{align*}
We say an $f^* \in \argmax_f \EV(f)$ is \emph{efficiency maximizing}.
Our goal is to find an $f$ that transforms the $z$ into the best
possible predictions in terms of efficiency.  Note that if it was
possible to predict exactly $p_i$ for ad $i$, these predictions would
maximize efficiency.

\paragraph{Questions} Ideally, we would like to use prediction maps
that are self-calibrated and efficiency-maximizing; we say such
prediction maps are \emph{nice}, and say a problem instance is
\emph{nice} if such a map exists.

First, we consider questions relating to the offline problem where we
have access to all the problem data.  Note that there must exist an
efficiency-maximizing prediction map.\footnote{Note $\EV$ depends only
  on the ordering of the ads for each query induced by $f$, and so
  over all possible $f$, $\EV$ takes on only a finite number of
  distinct values.}
\begin{description} \itemsep -1pt
\item[\Qnice] Are all problem instances nice? That is, do
  self-calibrated efficiency-maximizing prediction maps always exist?
\item[\Qpoly] Can an efficiency-maximizing prediction map, even one
  that is not self-calibrated, be found in polynomial time?
\end{description}

In practice, we are further concerned with learning a good prediction
map from observed data.  Suppose we start with some $f_0$, for example
the function that gives the predictions of the underlying system.
Then, we serve some large number of queries with this $f_0$, and
observe the results.  We would like to then train an improved $f_1$
from this data, serve another large batch of queries ranked using
$f_1$, then train an $f_2$, etc.

A natural procedure is to choose $f_t$ so that the predictions on the
ads shown in batch $t-1$ would have been calibrated under $f_t$.  Of
course, when we then select ads using $f_t$ on the next batch, we may
show different ads.  Formally, define $T: [0,1]^K \rightarrow [0,1]^K$
(a function from prediction maps to prediction maps) by $T(f) = f'$
where
\[ 
  f'(z) = 
\begin{cases}
\E_{f} [ p \mmid z]  & \text{when $\Pr_f(z) > 0$} \\
f(z) & \text{otherwise}.
\end{cases}
\]
We assume we have enough data in each batch so that we can calculate
$\E_{f_{t-1}} [ p \mmid z]$ exactly.  Then, we ask:
\begin{description}
\item[\Qfix] Does $T$ always have at most a small (polynomial) number
  of fixed points? 
\item[\Qconv] Does $T$ always have at least one fixed point where ads
  are shown? 
\end{description}
\Qfix is important, because with an affirmative answer we could
potentially enumerate the fixed points and find the best one from an
efficiency perspective.  A negative answer to \Qconv implies the
iterative calibration procedure will cycle.  
To see this, note that for a given starting point $f_0$, subsequent
$f_t(z)$ can only take on finitely many values: $\E[p \mmid z]$ for
some distribution of ads that show (finitely many values), or
$f_0(z)$. That means that $T$ maps some finite set of calibration maps
into itself. Since it has no fixed points, $T$ is a permutation and so
must cycle.

In the next two sections, we address these questions in the general
case (putting no additional restrictions on the problem instances).

\section{Mechanism \all: Threshold Selection}
\label{sec:all}
In this section, we consider the case where we select ads by mechanism
\all, that is, we show all ads where $b \cdot f(z) -1 \ge 0$.

We will show that an efficiency-maximizing prediction map can be found
efficiently (\Qpoly), but without further assumptions, \Qnice, \Qfix,
and \Qconv are answered in the negative.  We prove the negative results
first; for this purpose, it is sufficient to construct
counter-examples.

In this section, the examples we construct all require only a single
query where all of the candidates have the same raw prediction $z$.
Thus, choosing prediction map reduces to choosing a single value $\ph
\in [0, 1]$.  The selection rule simply shows all candidates where $b
\cdot f(z) = b \cdot \ph \ge 1$.

\newcommand{\T}{\rule{0pt}{2.2ex}}
\begin{figure*}[t!]
\begin{center}
  \begin{tabular}{|crrrr|r|}
    \hline
    Ad & CTR & bid & min $\ph$ & EV & cumulative CTR\\
    \hline
    1 &0.1 & $1 / (0.1) = 10.0$                 & 0.10 & 0.00 & 0.10  \\
    2 & 0.2 & $2 / (0.1 + 0.2) \approx\  6.7$   & 0.15 & 0.33 & 0.15  \\
    3 & 0.3 & $3 / (0.1 + 0.2 + 0.3) =\  5.0$   & 0.20 & 0.50 & 0.20  \\
    4 & 0.4 & $4 / (0.1 + \dots + 0.4) =\  4.0$ & 0.25 & 0.60 & 0.25  \\
    \hline
  \end{tabular}
\end{center}
\caption{An example with 5 fixed points, one for each prefix of the
  list of ads.  For each $i$, setting $\ph$ to the value in the ``min
  $\ph$'' column induces a fixed point where ads $1, \dots, i$ show. The
  fifth fixed point is the degenerate one that shows no ads, with say
  $\ph = 0$.}\label{fig:many}
\end{figure*}

\paragraph{\Qnice: All fixed points can have bad efficiency}
Consider an example with $2n + 1$ candidate ads, divided into three
classes, with ads given as $(p, b)$ tuples:
\begin{enumerate}[A)] \itemsep -2pt
\item  $1$ ad is   $(0.5, 2.0)$, shown if $\ph \ge 0.5$
\item  $n$ ads are $(1, 1.9)$, shown if $\ph \ge 1/1.9 \approx 0.53$
\item  $n$ ads are $(0, 1.8)$, shown if $\ph \ge 1/1.8 \approx 0.56$
\end{enumerate}
We either show no ads, $A$, $A \!+\! B$, or $A \!+\! B \!+\! C$.
Choosing $\ph = 0.5$ is a fixed point (it only shows the first ad)
which generates value $0.5 \cdot 2 - 1 = 0$.  Using $\ph = 0.54$ shows
$A + B$, and generates value $0.9n$.  But, this is not a fixed point:
the observed CTR is near one (for large $n$).  Showing all the ads
(which occurs for any $\ph > 1/1.8$) is not a fixed point, and
generates negative value, since ads from class $C$ generate value
$-n$.

\paragraph{\Qfix: An example with exponentially many fixed points}
Suppose there are $n$ candidates $(p_i, b_i)$ where the $p_i$ are
distinct, and we have indexed by $i$ so that $p_i$ is strictly
increasing.  Further, suppose $b_i = \frac{i}{p_{1:i}}$, a decreasing
sequence (using the shorthand $p_{1:i} \equiv \sum_{j=1}^i p_j$).
Pick any $i \in \set{1, \dots, n}$, and let $\ph = \frac{1}{b_i}$.  We
show candidate $j$ if $b_j \ph = \frac{b_j}{b_i} \geq 1$.  Since the
bids are decreasing, we show candidate $j$ if and only if $j \le i$.
Thus, serving with $\ph = \frac{1}{b_i} =\frac{p_{1:i}}{i}$ we show
candidates $1, \dots, i$, and so the average CTR is in fact $\ph$.
Thus, for any $i \in \set{1, \dots, n}$, there is a fixed-point $\ph$
that shows ads $\{1, \dots, i\}$.  Figure~\ref{fig:many} shows an
example of this construction.  If we have $m$ queries each with a
distinct fixed raw prediction $z$ and $n$ candidates constructed in
this manner, we can choose a per-query fixed point independently for
each query, for $n^m$ distinct fixed points.

\paragraph{\Qconv: An example with no fixed points}
Consider a single query with two candidates, $(p_1 = 0.7, b_1 = 4, z)$
and $(p_2=0.1, b_2 = 2, z)$.  For any $\ph \geq 0.5$, both ads show
and we observe a click-through-rate of $0.4$, so no such $\ph$ can be
self-calibrated.  For any $\ph \in [0.25, 0.5)$, only ad 1 shows, and
we observe a click-through rate of $0.7$.  For $\ph \in [0, 0.25)$, we
don't show any ads.  Thus, there is no non-trivial fixed point;
assuming we start with $\ph \ge 0.25$, the calibration procedure will
cycle between $0.7$ and $0.4$.

\paragraph{\Qpoly: Calculating the efficiency-maximizing $f$}
The above examples show that self-calibrated prediction maps may not
exist, and that even if they do, they need not maximize efficiency.

Nevertheless, given access to the full problem data (including true
click-through rates) one might be interested in calculating an
efficiency maximizing prediction map.  The following algorithm
accomplishes this in polynomial time.

We define $f^*$ by considering each $z' \in \set{1, 2, \dots, K}$
independently:
\begin{enumerate}
\item Consider the set of candidates $(p, b, z, q)$ where $z = z'$, and
  sort these candidates in decreasing order of bid, for $j=1, \dots,
  n_j$.  We must show some prefix of this list.  In
  particular, if we set $\ph = 1/b_j$ and $b_{j+1} < b_j$, then we
  will show exactly ads $1, \dots, j$.
\item For each $j$ where $b_{j+1} < b_j$, compute the expected value
  per query of using $\ph_j = 1/b_j$ (which shows ads 1, \dots, j).
  This can be computed as
  \[ \EV(\ph_j) = \sum_{i=1}^j \PQ(q_i) (p_i \cdot b_i - 1).\]
\item Let $f(z) = \ph_{j^*}$ where $\ph_{j^*}$ is the value that
  maximizes $\EV(\ph_j)$.
\end{enumerate}

While this result is interesting theoretically (especially in contrast
to results in the next section), we note it is not likely to be useful
in practice: if it was possible to estimate $p_i$ accurately for each
ad, then one could simply throw out the coarser-grained predictions
$z_i$ and use these estimates.

\section{Mechanism \one: Selecting One Ad}
\label{sec:one}
In this section, we consider results for selection mechanism \one.
When there is only a single query, or only a single raw prediction,
selection mechanism \one can be quickly analyzed, and our questions
are in fact answered in the affirmative, except for \Qfix. But in non-trivial
cases, we again show negative answers to all four questions.

\paragraph{Single query, multiple raw predictions}
Selection mechanism \one becomes rather degenerate under a single
query.  We show how to construct a nice $f$, answering \Qnice and
\Qpoly, and \Qconv in the affirmative.

For each raw prediction $z' \in \set{1, \dots, K}$, observe that if an
ad with $z_i = z'$ shows, it must be an ad that has bid $b(z') \equiv
\max_{j:z_j = z'} b_j$.  Thus, if an ad with $z'$ shows, the expected
value generated is $b(z') \cdot \EC[p \mmid z', b(z')]$, where $\EC[p
\mid z', b(z')]$ is the average click-through-rate of ads with $z =
z', b = b(z')$.  We can guarantee we obtain this value by simply
setting $f(z') = \EC[p \mmid z', b(z')]$ and $f(z) = 0$ for all $z
\neq z'$.  Note that this $f$ is self-calibrated because ties are
broken uniformly at random under selection mechanism \one, answering
\Qconv in the affirmative.
We obtain maximum efficiency by using the $f$ that only shows ads with
raw prediction
\[z^* = \argmax_z b(z) \cdot \EC[p \mmid z, b(z)].\]

Let $f_z$ be the $f$ function that only shows candidates with the
given $z$ value.  Thus, $f_{z^*}$ is nice.  However, we can define a
more satisfying $f^*$ by 
\[ %
f^*(z) = \E_{f_z}[p \mmid z].
\]
We only show ads $(b, z)$ where $b \cdot f^*(z)$ achieves the
argmax value over candidates, and in fact 
\[
 b \cdot f^*(z) = b(z) \cdot \E_{f_z}[p \mmid z],
\]
and so we still maximize efficiency.

\newcommand{\IC}{\mathcal{I}} 
The answer to \Qfix is negative: iterative calibration can have
exponentially many fixed points. Suppose each ad $i$ has a distinct
$z_i$, and $p_i = b_i ^ {-1}$.  Let $\IC$ be any subset of the ads and
define $f_\IC$: $f_\IC(z_i) = p_i$ for $i \in \IC$, $f_\IC(z_i) = 0$
for $i \not\in \IC$. Then, under $f_\IC$ all ads in $\IC$ tie, so we
show them randomly. Each of the $2^{\abs{\CS}}$ subsets of $\CS$ thus
corresponds to a self-calibrated prediction map that shows a different
set of ads.

\paragraph{Multiple queries, single raw prediction}
Under mechanism \one, if there is a single raw prediction $z$ made for
all candidates (on all queries), then the ads that show are in fact
independent of the value $\ph = f(z) > 0$: for each query, we always
randomly pick one of the candidates with the highest bid.  Thus, any
$\ph > 0$ is efficiency-maximizing, and we can choose $\ph$ equal to the
average observed CTR to obtain self-calibration.  Thus, in this case
we answer \Qnice - \Qconv in the affirmative.

\paragraph{\Qpoly: NP-hardness in general}
In general (with at least two distinct raw predictions and at least
two queries), under selection mechanism \one, the offline problem of
finding the efficiency-maximizing prediction map $f$ is NP-hard, even
if all bids are $1$.  We show this using a reduction from the minimum
feedback arc set (MFAS) problem on tournaments (see, for example,
\citet{kleinberg10sleeping}).

In this problem, there are $n$ players, $\set{1, \dots, n}$, that have
just completed a tournament where every pair of players has played.
The MFAS for this problem is a ranking of the players that minimizes
the number of upsets; that is, if $\mu_i$ is the rank of player $i$,
we want a ranking $\mu$ that minimizes the number of times $\mu_i >
\mu_j$, but player $j$ beat player $i$.

We encode this problem as an auction efficiency maximization problem
as follows: There are $n$ distinct $z$ values, $1, \dots, n$, one for
each player, and there are $\h n(n-1)$ queries (each equally likely),
one for each $(i, j)$ pair with $i < j$.
The query for the pair $(i,j)$ (where $i$ beat $j$ without loss of
generality) has two candidates $(p, z)$, namely $(1, i)$ and $(0, j)$.
Thus, if we show the ad corresponding to the winner (with $z = i$), we
have $p=1$, and the bid is $1$, so we get value 1; if we show ad with
$z=j$, we have $p=0$, we get no value.  It is then clear that the
efficiency-maximizing ranking of the raw predictions $z$ exactly
corresponds to the solution to the MFAS problem.

\paragraph{ Negative results for \Qnice, \Qfix, and
  \Qconv in general}
We also show negative results for \Qnice, \Qfix, and \Qconv in
general.

For \Qnice, observe that in the NP-hardness construction when there is a
perfect ranking, we observe a CTR of 1.0, and so the
efficiency-maximizing prediction map cannot be self-calibrated.  We
can illustrate this directly with the following example.  There are
four ads, each given as $(p, b, z)$ tuples:

\begin{center}
\begin{tabular}{ll}
\ch{$q_1$} & \ch{$q_2$} \\
\hline
A $(1.0, 2, z_1)$ & C $(1.0, 2, z_2 )$ \\
B $(0.0, 2, z_2)$ & D $(0.0, 1, z_1)$
\end{tabular}\end{center}

\noindent
We need $f(z_1) > f(z_2)$ in order to guarantee we show Ad A on $q_1$;
we also need $f(z_2) > \h f(z_1)$ in order to show Ad C on $q_2$.  We
will observe a 1.0 CTR on both $z_1$ and $z_2$ on any such efficiency
maximizing $f$, but we are constrained to pick $f(z_2) < f(z_1) \le
1$, and so no such $f$ can be self-calibrated.

For \Qfix, we have already shown multiple fixed points in the
single-query case.  If we consider multiple queries, where each query
has a single distinct raw prediction, we immediately arrive at a
problem with exponentially many fixed points.

For \Qconv, it is straightforward to construct an example with cycles,
but constructing one with no fixed point is a bit trickier.  In
particular, any time there is some prediction $z$ where each query has
at least one ad with prediction $z$, we can always find a fixed point
by setting $f(z') = 0$ for $z' \ne z$ and $f(z) > 0$.  The set of ads
shown will be independent of the non-zero value $f(z)$, so we can set
it equal to the observed CTR, achieving self-calibration (except in
the degenerate case where all the ads with prediction $z$ have zero
CTR).

However, it is still possible to construct problems with no fixed
points without resorting to such degeneracy, as the following example
illustrates.  Each query is equally likely, all the bids are $1$, and
the $(p,z)$ ad tuples are: \\

\vspace{-0.1in}
\begin{center}
\begin{tabular}{llll}
\ch{$q_1$} & \ch{$q_2$} & \ch{$q_1$} & \ch{$q_2$} \\
\hline
A $(0.5, z_1)$ & B $(0.6, z_2)$ & C $(0.5, z_1)$ & E $(0.2, z_2)$ \\
               &                & D $(0.6, z_2)$ & F $(0.3, z_1)$ 
\end{tabular}\end{center}
If $f(z_1) > f(z_2)$, then we show ads A,B, C, and F.  In this case,
we observe a CTR of $(0.5 + 0.5 + 0.3)/3 = 0.433$ for $z_1$, and 0.6
for $z_2$, so we cannot be self-calibrated.
If $f(z_1) < f(z_2)$, we show ads A, B, D, and E, and observe a CTR of
$(0.6 + 0.6 + 0.2)/3 = 0.467$ for $z_2$, and $0.5$ for $z_1$, and so
again we cannot be self-calibrated.
Finally, if $f(z_1) = f(z_2)$, we always show $A$ and $B$, and show
the other ads half of the time.  Thus, we observe a CTR of $(3/4) 0.5
+ (1/4) 0.3 = 0.45$ for $z_1$, and a CTR of $(3/4) 0.6 + (1/4) 0.2 =
0.5$ for $z_2$, and so again we cannot be well-calibrated.  Thus, no
self-calibrated $f$ exists for this problem.

\section{Sufficient Conditions}\label{sec:cond}

\newcommand{\zbqstrong}{Prop \texttt{E1}\xspace}
\newcommand{\zbonly}{Prop \texttt{E2}\xspace}

\newcommand{\si}{Prop \texttt{SI}\xspace}

\begin{table}[t!]
\begin{center}
\begin{tabular}{|l r@{} c@{} l l|}
\hline
both & \zbqstrong &\imp& \zbonly & (immediate) \\
\all & \zbonly &$\iff$& \si & Thm~\ref{thm:equivall}\\
\hline
\all & \zbonly &\imp& nice  & Thm~\ref{thm:e2ev} \\
\all & \zbqstrong &\imp& nice & (from above) \\
\one & \zbqstrong &\imp& nice & Thm~\ref{thm:onezgqev} \\
\hline
both & \si &\nimp& \zbqstrong & Sec~\ref{sec:negresults} \\
\one & \zbonly &\nimp& \si    & Sec~\ref{sec:negresults} \\
\one & \si &\nimp& nice       & Sec~\ref{sec:negresults} \\
\hline
\end{tabular}
\end{center}
\caption{
  Relationships between problem properties.  A ``nice'' problem instance 
  is one where a self-calibrated efficiency-maximizing prediction map exists.
}
  \label{tab:results}
\end{table}

As the previous two sections show, without additional assumptions
significant problems arise if one tries to achieve both calibration
and auction efficiency.  In this section, we introduce additional
assumptions that are sufficient to guarantee nice prediction maps
exist. Table~\ref{tab:results} summarizes our results.

The intuition behind our results is a basic property of
conditional probability. Calibration depends on the conditional
expectation $\E[p \mmid z]$. In general, selection changes the
distribution this expectation is with respect to. But if selection
is \emph{only} a function of $z$, it does not change the conditional
distribution of $p$ given $z$, since the latter is already conditioned
on $z$.

For example, suppose we have a single query, and that all bids are 1,
so all selection decisions are functions of $z$. This means that $\E[p
\mmid z]$ does not change under selection, and thus defines an
efficiency-maximizing self-calibrated prediction map.  To extend this
intuition to more realistic auctions, we need to make sure that the
query and the bid do not add any information about $p$, so that
selection does not change $\E[p \mmid z]$ and the different $\E[p
\mmid z]$ for each query can be reconciled.
We now state these properties formally:

\paragraph{\zbqstrong}  For each $z$ there exists a value $\pb(z)$ such
that for each query $q$ with $\Pr_\CS(q \mmid z) > 0$, and for each
$b$ with $\Pr_\CS(b \mmid q, z) > 0$,
\begin{equation}\label{eq:pzbindep}
  \EC[p \mmid z,b,q] = \EC[p \mmid z,q] = \EC[p \mmid z] \equiv \pb(z).
\end{equation}
That is, in all cases the bid and query provide no more information
than the raw prediction about average click-through
rates.\footnote{Note that this does not hold under the NP-Hardness
  reduction for \one in the previous section, as $\EC[p \mmid z, q]
  \neq \EC[p \mmid z]$.}  For this assumption, the natural prediction
map to consider is $f(z) = \pb(z)$. \qed

\paragraph{\zbonly} A weaker assumption is that
\begin{equation}
  \EC[p \mmid z,b] = \EC[p \mmid z]
\end{equation}
whenever both expectations are defined.  This essentially marginalizes
over queries, rather than holding simultaneously for all $q$.\qed

\paragraph{\si} A problem instance is \textbf{selection-invariant} if
for all $f, f'$, for any $z$ where both $\E_f[p \mmid z]$ and
$\E_{f'}[p \mmid z]$ are defined, we have
\begin{equation}
  \E_f[p \mmid z] = \E_{f'}[p \mmid z].
\end{equation}
Selection invariance says that the observed CTR for a given raw
prediction $z$ is independent of the prediction map used for selection.
Under this assumption, the natural calibration map to consider is
$
f^*(z) = \E_{f_z}[p \mmid z],
$
where $f_z$ is any prediction map that shows some ads with raw
prediction $z$. \qed

It is easy to show that \zbqstrong implies \zbonly.

A weak per-query variant of \zbqstrong is that, for all $z$, $b$, and
$q$ (when defined), $\EC[p \mmid z,b,q] = \EC[p \mmid z, q]$.  We can
dismiss this assumption as insufficient, as we can take the negative
examples of Section~\ref{sec:all} and re-state them where each
candidate occurs on a distinct query, each equally likely.  Thus, the
above property holds trivially, but the pathological behaviors still
occur.

\subsection{Properties that Imply Nice Maps Exist}
First, we show that under mechanism \all, \zbonly and \si are
equivalent; we then show that \zbonly (and hence also \si) imply a
nice problem.

\begin{theorem}\label{thm:equivall}
  Under selection mechanism \all, \zbonly is equivalent to \si (selection
  invariance).
\end{theorem} 

\begin{proof}[Proof sketch]
  Suppose \zbonly holds.  Selection mechanism \all must show either
  all of the candidates with a given $(z, b)$ combination, or none of
  them.  Thus, for any $f$ where $\Pr_f(z, b) > 0$, we must have 
  \begin{equation}\label{eq:sa}
  \E_f[p \mmid z,b] = \EC[p \mmid z,b].
  \end{equation}
  Then, for any $f$, assuming $\E_f[p \mmid z]$ is defined,
  \begin{align*}
    \E_f[p \mmid z] 
    &= \E_f[\E_f[p \mmid z,b]] \\
    &= \E_f[\EC[p \mmid z,b]] && \text{\eqr{sa}} \\
    &= \E_f[\EC[p \mmid z]] && \text{\zbonly} \\
    &= \EC[p \mmid z].
  \end{align*}
  For the other direction, suppose we have selection invariance (\si).
  It is sufficient to consider a fixed raw prediction $z$ (if there
  are multiple $z$, we can consider them independently). Also,
  we can assume candidates have distinct bids - if multiple candidates
  have the same bid and raw prediction, mechanism \all treats them all
  the same, so we can just average over them.

  Index the bids $(b_1, b_2, \dots)$ in decreasing order. Then, depending
  on the chosen $\ph = f(z)$, we either show (when the appropriate queries
  occur) ad $1$, or ads $1$ and $2$, etc. \si says that no matter what
  $\ph$ is, the average CTR of the ads we show is the same.
  Suppose that all the ads are on the same query. Then \si implies
  $p_1$ = $\h p_1 + \h p_2$, so $p_1 = p_2$;
  $\h (p_1 + p2) = \frac{1}{3} (p_1 + p_2 + p_3)$, so $p_1 = p_2 = p_3$; and
  so on. When the ads are on different queries, the weights in the above
  equalities change to reflect the query distribution, but are still all
  positive and sum to 1, so the same inductive reasoning holds.
\end{proof}
This result implies that under selection mechanism \all, when \zbonly
holds the prediction map $f^*(z) = \EC[p \mmid z]$ is self-calibrated.
Next, we show this map is in fact also efficiency-maximizing:

\begin{theorem}\label{thm:e2ev}
  Under selection mechanism \all, \zbonly implies $f^*$ is efficiency
  maximizing, where $f^*(z) = \EC[p \mmid z]$.
\end{theorem} 

\begin{proof}
Recall we need to show $f^*$ maximizes
\[
  EV(f) = \sum_{i \in \CS} \PQ(q_i) \Pr(i \mmid q_i, f) (p_i b_i - 1).
\]
Since selection decisions for one $z$ value do not impact others, it
suffices to consider a single $z$ value.  We can decompose the sum
over $\CS$ over the partition that associates all the ads that share a
common bid and raw prediction.  Let $B = \set{i \mmid b_i = b, z_i =
  z} \subseteq \CS$ be the element of this partition for $(b,z)$.  For
a given $f(z) = \ph$, either all the ads in $B$ show (when their
respective queries occur), or none of them do; thus, if we can show
that $f^*$ shows these ads if and only if they increase EV, we are
done.  The expected value per query of showing these ads is:
\begin{equation}
  \sum_{i \in B} \PQ(q_i) \Pr(i \mmid q_i, f) (p_i b_i - 1).
\end{equation}
Since $\Pr(i \mmid q_i, f) \in \set{0, 1}$ must be the same for all
these ads, this quantity is non-negative if and only if $\sum_{i
  \in B} \PQ(q_i) (p_i b_i - 1) \geq 0$.

Recall $\PrC(i) = \PQ(q_i)/C$ where $C = \sum_{i \in \CS} \PQ(q_i)$.
We have $\PrC(i \wedge b \wedge z) = \PrC(i)$ if $i \in B$, and 0 otherwise.
Letting $C_B = \sum_{i \in B} \PQ(q_i)$, then $\PrC(b \wedge z) =
\frac{C_B}{C}$, and so
\begin{equation}\label{eq:ii}
\PrC(i \mmid b, z) 
= \frac{\PrC(i)}{C_B/C} 
= \frac{\PQ(q_i)/C}{C_B/C} 
= \frac{\PQ(q_i)}{C_B}
\end{equation}
for $i \in B$, and 0 otherwise.
Then,
\begin{align*}
\EC[p \mmid z] 
  &= \EC[p \mmid b, z] && \text{\zbonly}\\
  &= \sum_{i \in \CS} \PrC(i \mmid b, z) p_i \\
  &= \frac{1}{C_B}\sum_{i \in B} \PQ(q_i) p_i.  && \text{\eqr{ii}}
\end{align*}
Using this result, we have
\begin{align*}
\sum_{i \in B} \PQ(q_i) (p_i b_i - 1) 
  &= b \left(\sum_{i \in B} \PQ(q_i) p_i\right)  - C_B) \\
  &= C_B (b \EC[p \mmid z] - 1).
\end{align*}
This quantity is non-negative if and only if $b f^*(z) - 1 \geq 0$;
since this is exactly the condition we use to decide whether or not to
show the ads in $B$, we are done.
\end{proof}

It is not hard to directly prove that under selection mechanism \all, \si
implies $f^*$ is efficiency-maximizing: the idea is to consider again
a single $z$, sort the ads by bid into blocks, and show by induction
that each block has average CTR $f^*(z)$.

In Section~\ref{sec:one} we saw that the problem of finding an
efficiency-maximizing $f$ is NP-hard under mechanism \one, even under
the assumption of a single bid.  Under \zbqstrong, fortunately the
situation is much easier:

\begin{theorem}\label{thm:onezgqev}
  Under selection mechanism \one, if \zbqstrong holds then the
  prediction map $f^*$ where $f^*(z) = \EC[p \mmid z]$ is
  efficiency-maximizing and self-calibrated.
\end{theorem}

\newcommand{\BP}{\mathfrak{B}^q}
\newcommand{\B}{B^q_{b,z}}

\begin{proof}
  For a query $q$, consider a partition $\BP$ of $\CS(q)$ into sets of ads
  that share a common $b$ and $z$, so the elements of the partition are
  \[
   \B = \set{i \mmid b_i = b, z_i = z, q_i = q} \subseteq \CS(q)
  \]
  for each $(b,z)$ pair.

  All $i \in B$ for some $B$ must share a common value $\Pr_f(i \mmid
  q)$.  We also use $B$ as the event that some $i \in B$ shows; so for
  example $\Pr_f(B \mmid q)$ is the probability that \emph{some} ad
  from $B$ shows.  Under selection mechanism \one, for each $i \in B$,
  we have $\Pr_f(i \mmid B, q) = \frac{1}{\abs{B}}$ (since ties are
  broken at random).
  Also,
  \begin{equation}\label{eq:ECpz}
    \EC[p \mmid b, z, q] = \frac{1}{\abs{\B}} \sum_{i \in \B} p_i.
  \end{equation}
  Recalling cost is zero under \one, for any $f$, 
  \begin{align*}
    &\EV(f) \\
    &= \sum_{q \in \Q} \PQ(q) \sum_{i \in C(q)} \Pr_f(i \mmid q) p_i b_i\\
    &= \sum_{q \in \Q} \PQ(q) \sum_{\B \in \BP} 
         \sum_{i \in \B} \Pr_f(i \mmid q) p_i b_i \\
    &= \sum_{q \in \Q} \PQ(q) \sum_{\B \in \BP} 
         \Pr_f(\B \mmid q) \frac{1}{\abs{\B}} \sum_{i \in \B}  p_i b  \\
\intertext{and using \eqr{ECpz},}
    &= \sum_{q \in \Q} \PQ(q) \sum_{\B \in \BP} 
         \Pr_f(\B \mmid q)  b\, \EC[p \mmid b, z, q] \\
    &\leq \sum_{q \in \Q} \PQ(q) \max_{\B \in \BP} 
         b\, \EC[p \mmid b, z, q].
  \end{align*}
  Thus, it is sufficient to show that selecting ads using $f^*$
  produces the expected value in the last line of the above
  inequality.  For each query, we rank the ads using $b \cdot f^*(z) =
  b\, \EC[p \mmid b, z, q]$, and so this is exactly the expected value
  that $f^*$ obtains.

  To see that $f^*$ is self-calibrated, observe that when 
  $\Pr_f(z, b, q) > 0$, 
  \[
   \E_f[p \mmid z, b, q] = \EC[p \mmid z, b, q] = f^*(z),
  \]
  and so
  \[
  \E_f[p \mmid z] 
    = \sum_{b,q} \Pr_f(b, q \mmid z) \E_f[p \mmid z, b, q] 
    = f^*(z).
  \]
  \qed
\end{proof}

\subsection{Negative Results}
\label{sec:negresults}
We show several negative results relating to the assumptions
considered in the previous section.

\paragraph{\one and \all: \si does not imply \zbqstrong}
Consider an example with two queries, each equally likely.  Each query
has two candidates, given as the following $(p, b)$ tuples (they all
share a common $z$):
\begin{center}
\begin{tabular}{ll}
\ch{$q_1$} & \ch{$q_2$} \\
\hline
A (0.1, 1) & C (0.1, 2) \\
B (0.2, 2) & D (0.2, 1)
\end{tabular}\end{center}
Because of the symmetry between these queries, under any $f$ (and
either selection mechanism), ad A must show with the same probability
as ad D, as must ads B and C.  Thus, for any $f$, $\E_f[p \mmid b=1, z]
= 0.15$, and similarly $\E_f[p \mmid b=2, z] = 0.15$.  Thus, selection
invariance holds, as does \zbonly.  However, $\EC[p \mmid z, b=1, q_1]
= 0.1 \neq \EC[p \mmid z, q_1] = 0.15$.

\paragraph{\one: \zbonly does not imply \si}
Consider the example, with two equally likely queries, and two
distinct raw predictions:
\begin{center}
\begin{tabular}{ll}
\ch{$q_1$} & \ch{$q_2$} \\
\hline
A $(0.2, 2, z_1$)  & C $(0.1, 2, z_1)$ \\
B $(0.1, 1, z_1)$  & D $(0.2, 1, z_1)$ \\
E $(1.0, 9, z_2)$ & 
\end{tabular}
\end{center}
Note that $\EC[p \mmid z_1, b=1] = \EC[p \mmid z_1, b=2] = 0.15$.  However,
if we consider two prediction maps $f(z_1) = 0.5, f(z_2) = 1$ and $f'(z_1)
= 1, f'(z_2) = 0$, under selection mechanism \one, we have $\E_f[p \mmid z_1]
= 0.1$, but $\E_{f'}[p \mmid z_1] = 0.15$.

\paragraph{\one: \si does not imply a nice problem}
We have four queries, each equally likely; the bids for the ads on
$q_3$ and $q_4$ are defined in terms of some small $\eps > 0$, with
$(p,b,z)$ tuples:

\newcommand{\tx}[1]{$(#1)\!\!$}
\begin{center}
\begin{tabular}{llll}
\ch{$q_1$}      & \ch{$q_2$}       & \ch{$q_3$}          & \ch{$q_4$} \\
\hline
A \tx{1,2,z_1} & C \tx{1,2,z_2} 
  & A' \tx{0,2\eps,z_1} & C' \tx{0,2\eps,z_2} \\
B \tx{0,2,z_2} & D \tx{0,1,z_1}  
  &  B' \tx{1,2\eps,z_2} & D' \tx{1,1\eps,z_1}  \\
\end{tabular}
\end{center}

Note that $q_3$ and $q_4$ mirror $q_1$ and $q_2$, except that the bids
are scaled by $\eps$, and the CTRs are reversed.  Under any $f$, ads
$A$ and $A'$ show with the same probability, as do $B$ and $B'$, and
the other two pairs.  Thus, under selection by any $f$, we have $\E_f[p
\mmid z_1] = \E_f[p \mmid z_2]= 0.5$ whenever the expectation is
defined, and so \si holds.  However, as $\eps \rightarrow 0$, only
$q_1$ and $q_2$ have any impact on efficiency.  Thus, as before we
have constraints on the optimal solution that $f(z_1) > f(z_2) > \h
f(z_1)$.  Thus, the prediction map $f^*$ with $f^*(z_1) = 0.5$ and
$f^*(z_2) = 0.5$ is not efficiency-maximizing, as it only shows ad A
on $q_1$ only half the time.

\section{Discussion and Future Work}

Our sufficient conditions are quite strong, but not unrealistic. They
require that the bid and query not add any information about the CTR,
conditional on the raw prediction. CTR estimation systems normally use
queries as features (e.g., \citep{graepel10webscale}), so it is
reasonable to hope that the query does not add extra information. Bids
are set by advertisers for query-ad pairs, which are already used by
CTR estimation systems, so any systematic patterns in bids are likely
to be accounted for. Since advertisers have much less information than
the auctioneer, it seems unlikely that they can add extra information
about CTRs through fine-grained bid manipulation. We can test if our
sufficient conditions hold by running randomization experiments that
change the mix of ads shown.

Since randomized predictions cannot in general lead to maximum
efficiency, it is natural to first consider deterministic prediction
maps.  Nevertheless, given the negative results in the current work,
it would be interesting to also study randomized calibration
strategies that provide calibration guarantees without needing IID
assumptions.  Then the natural question becomes: how much efficiency
is lost by using a randomized calibration strategy, versus using a
deterministic efficiency-maximizing prediction map that is not
self-calibrated.

\bibliography{../new,../my_pubs,calibration_references}
\bibliographystyle{plainnat} 

\end{document}